\documentclass[aps,pre,twocolumn,eqsecnum,showpacs,superscriptaddress]{revtex4-1}
\usepackage{graphicx,color}
\usepackage{amsmath}
\usepackage{amssymb}
\usepackage{bm}

\renewcommand{\Re}{\mathop{\text{Re}}\nolimits}
\renewcommand{\Im}{\mathop{\text{Im}}\nolimits}

\newcommand{\Tr}{\mathop{\text{Tr}}\nolimits}
\newcommand{\ket}[1]{|{#1}\rangle}
\newcommand{\bra}[1]{\langle{#1}|}
\newcommand{\bras}[2]{{}_{#2}\hspace*{-0.2mm}\langle{#1}|}

\newcommand{\Var}{\mathop{\text{Var}}\nolimits}

\newcommand{\pv}{\mathop{\text{P}}\nolimits}

\definecolor{dgreen}{rgb}{0,0.5,0}

\definecolor{delete}{cmyk}{0.5,0,0,0}

\begin{document}

\title{Typical pure nonequilibrium steady states and irreversibility for quantum transports}



\author{Takaaki Monnai}
\affiliation{Department of Materials and Life Sciences, Seikei University, Tokyo 180-8633, Japan}
\author{Kazuya Yuasa}
\affiliation{Department of Physics, Waseda University, Tokyo 169-8555, Japan}



\date{\today}

\begin{abstract}
It is known that each single typical pure state in an energy shell of a large isolated quantum system well represents a thermal equilibrium state of the system.
We show that such typicality holds also for nonequilibrium steady states (NESS's).
We consider a small quantum system coupled to multiple infinite reservoirs.
In the long run, the total system reaches a unique NESS\@.
We identify a large Hilbert space from which pure states of the system are to be sampled randomly and show that the typical pure states
well describe the NESS\@.
We also point out that the irreversible relaxation to the unique NESS is important to the typicality of the pure NESS's.
\end{abstract}
\pacs{%
05.30.Ch, 
05.60.Gg, 
03.65.Yz, 
05.70.Ln
}


\maketitle

\section{Introduction}
Recently, considerable attention has been paid to the ``typicality'' in the exploration of the foundations of statistical mechanics.
It has been understood that each single typical pure state in an energy shell well describes a thermal equilibrium state of a large quantum system.
Indeed, a vast majority of the pure states in an energy shell yield expectation values very close to those evaluated with the microcanonical ensemble for a class of observables (microcanonical typicality) \cite{Sugita1b,*Sugita1,*Sugita1arXiv,Reimann1,Reimann2008-JSP}.
If we restrict ourselves to observables of a small subsystem, their expectation values in the typical pure states well coincide with those evaluated with the canonical ensemble for the subsystem (canonical typicality) \cite{ref:TasakiCanonicalTypicality,Lebowitz1,Popescu1-arXiv,*Popescu1} (see also \cite{ref:GemmerEPJB,ref:GemmerTextbook}).
It has also been argued that given a realistic Hamiltonian of interacting particles each energy eigenstate looks like a chaotic thermal state (eigenstate thermalization) \cite{ref:TasakiCanonicalTypicality,Rigol1,ETH-Reimann,ref:GiacomoVittorioETH}.
The importance of macro observables is discussed \cite{Lebowitz3} (see also \cite{ref:VanKampen,vonNeumann1,*vonNeumann1arXiv}).
Typicality allows us to compute expectation values and thermodynamical quantities at a finite temperature only with a single pure state \cite{Sugiura1,Sugiura2}.
It is also allowed to calculate higher-order moments of observables \cite{Monnai1}. 
This is important in evaluating physical quantities in the Heisenberg picture and thus in analyzing nonequilibrium processes starting from a thermal equilibrium state. 
Equilibration/thermalization \cite{Rigol1,Reimann2,Popescu2,Lebowitz2,Short1,Reimann3,Reimann4,Kaminishi:2015aa}, temporal fluctuations around equilibrium \cite{Popescu3}, and relaxation times \cite{Goldstein1,Monnai2,Goldstein1b} have also been subjects under intense study recently in the context of typicality, and there are beautiful experimental works on the relevant issues \cite{Kinoshita:2006aa,ref:Schmidmayer-AtomChip-NaturePhys9,GGE-exp}.

It is a challenging problem to explore the typicality for nonequilibrium systems.
In particular, the nonequilibrium steady states (NESS's) with finite stationary currents are of great interest \cite{ref:NESS-Antal,ref:NESS-Ruelle,Tasaki3,ref:NESS-TasakiMatsui,ref:NESS-AschbacherPillet-JSP,ref:NESS-AschbacherJaksicPautratPillet,ref:NESS-Tasaki,Gaspard1,Saito1,ref:SaitoTasaki-ExtendedClausius}.
In Ref.\ \cite{ref:TypicalNESS}, we considered a NESS realized in a setup consisting of two infinite reservoirs interacting locally with each other, and showed that there exists a large Hilbert space whose typical pure states well describe the NESS: the typicality holds also for NESS's.
We sample a typical pure state $\ket{\phi}$ randomly from a Hilbert space $\mathcal{H}_{E_1,E_2}=\mathcal{H}_{E_1}\otimes\mathcal{H}_{E_2}$, where $\mathcal{H}_{E_\nu}$ ($\nu=1,2$) is the Hilbert space representing an energy shell $[E_\nu,E_\nu+\Delta E]$ of the $\nu$th reservoir, and scatter it by a M\o ller wave operator $\hat{W}$ to construct a pure state $\ket{\phi}_\text{NESS}=\hat{W}\ket{\phi}$.
Such pure states $\ket{\phi}_\text{NESS}$ are typically equivalent to the NESS, in the sense that the expectation values of an observable $\hat{A}$ in the typical pure states $\ket{\phi}_\text{NESS}$ are very close to the expectation value in the NESS: $\bras{\phi}{\text{NESS}}\hat{A}\ket{\phi}_\text{NESS}\simeq\langle\hat{A}\rangle_\text{NESS}$.
We call the pure states $\ket{\phi}_\text{NESS}$ \textit{typical pure NESS's}.
We remark that the initial pure states $\ket{\phi}$ sampled from $\mathcal{H}_{E_1,E_2}$ are generally entangled states, rather than product states of the form $\ket{\phi_1}\otimes\ket{\phi_2}$ sampled separately from the Hilbert spaces $\mathcal{H}_{E_1}$ and $\mathcal{H}_{E_2}$.

In this article, we generalize this construction of the typical pure NESS's to the case where a small quantum system $S$ (e.g., a quantum dot) interacts with multiple reservoirs, often found in mesoscopic quantum junction setups.
We will see that the generalization is not trivial, due to the presence of the small system $S$.
First of all, it is not clear at first glance from which subspace, in particular of the small system $S$, typical pure states $\ket{\phi}$ to be scattered to construct typical pure NESS's $\ket{\phi}_\text{NESS}=\hat{W}\ket{\phi}$ should be sampled.
We are going to identify the relevant Hilbert space $\mathcal{H}_\text{NESS}$.
To this end, we will notice that the irreversible relaxation of the small system $S$ is important for the construction of the typical pure NESS's $\ket{\phi}_\text{NESS}$.
Moreover, we will point out that the naive perturbative approach is not useful to capture the irreversibility and to derive the NESS, in contrast to the case in the absence of the small system $S$\@.
In these respects, the present setup with the small system $S$ requires additional cares to study the typicality of pure NESS's $\ket{\phi}_\text{NESS}$, making the generalization nontrivial.

This article is organized as follows.
We start by briefly recapitulating the typicality for equilibrium systems in Sec.\ \ref{sec:TypicalEq} and the standard approach to describe NESS's in Sec.\ \ref{sec:NESS}\@.
We then provide the construction of pure NESS's $\ket{\phi}_\text{NESS}$ in the presence of a small system $S$ in Sec.\ \ref{sec:Construction} under the assumption of irreversible relaxation to a unique NESS, and prove their typicality in the relevant Hilbert space $\mathcal{H}_\text{NESS}$ in Sec.\ \ref{sec:TypicalNESS}\@.
To illustrate the relevance of the assumptions, we look at an exactly solvable model in Sec.\ \ref{sec:Model}.
The article is finally summarized in Sec.\ \ref{sec:Summary}\@.

\section{Typicality for Equilibrium Systems}
\label{sec:TypicalEq}
Before starting to discuss the typicality for NESS's, let us briefly recapitulate the typicality for equilibrium systems.
Here, we focus on the ``microcanonical typicality'' \cite{Sugita1b,*Sugita1,*Sugita1arXiv,Reimann1,Reimann2008-JSP,Sugiura1} (see \cite{ref:TasakiCanonicalTypicality,ref:GemmerEPJB,Lebowitz1,Popescu1-arXiv,*Popescu1,Sugiura2} for the ``canonical typicality'').

We consider a large quantum system and pick a pure state $\ket{\psi}$ randomly from the Hilbert space $\mathcal{H}_E$ spanned by the energy eigenstates $\ket{E_j}$ belonging to the energies $E_j$ within an energy shell $[E,E+\Delta E]$ (with the number of particles fixed at $N$).
The pure state $\ket{\psi}$ is given by
\begin{equation}
\ket{\psi}
=\sum_{j=1}^dc_j\ket{E_j},
\end{equation}
where $d=\dim\mathcal{H}_E$ is the dimension of the Hilbert space $\mathcal{H}_E$.
We sample the pure states $\ket{\psi}$ uniformly from $\mathcal{H}_E$ according to the Haar measure given by
\begin{equation}
d\mu(\ket{\psi})
\propto
\delta\!\left(
\sum_{j=1}^d
|c_j|^2-1
\right)
\prod_{j=1}^dd^2c_j.
\label{eqn:HaarEq}
\end{equation}
Recalling the formulas 
\begin{equation}
\overline{|c_j|^2}=\frac{1}{d},\quad
\overline{|c_j|^2|c_{j'}|^2}
=\frac{1}{d(d+1)}
(1+\delta_{jj'}),
\label{eqn:Cmoments}
\end{equation}
with the others up to the fourth moments of $c_j$ vanishing \cite{Ullah196465,ref:GemmerTextbook}
[where $\overline{\mathcal{O}}=\int d\mu(\ket{\psi})\mathcal{O}$ denotes the ensemble average over the Haar measure], the average and the variance of the expectation value $\bra{\psi}\hat{A}\ket{\psi}$ of an observable $\hat{A}$ over the uniformly sampled states $\ket{\psi}$ are calculated as 
\begin{subequations}
\label{eqn:AveVarEq}
\begin{equation}
\overline{\bra{\psi}\hat{A}\ket{\psi}}
=\frac{1}{d}\sum_{j=1}^d\bra{E_j}\hat{A}\ket{E_j}=\langle\hat{A}\rangle_\text{mc},
\label{eqn:AveEq}
\end{equation}
and
\begin{align}
\Var[\bra{\psi}\hat{A}\ket{\psi}]
={}&
\frac{1}{d(d+1)}\sum_{j=1}^d\sum_{j'=1}^d|\bra{E_j}\hat{A}\ket{E_{j'}}|^2
\nonumber\\
&{}-\frac{1}{d^2(d+1)}\left(
\sum_{j=1}^d\bra{E_j}\hat{A}\ket{E_j}
\right)^2
\nonumber\\
\le{}&\frac{(\Delta\hat{A})^2_\text{mc}}{d+1},
\label{eqn:VarEq}
\end{align}
\end{subequations}
respectively, where $\Var[\mathcal{O}]=\overline{\mathcal{O}^2}-\overline{\mathcal{O}}^2$ is the ensemble variance over the Haar measure, while $\langle\hat{A}\rangle_\text{mc}=\Tr\{\hat{\rho}_\text{mc}\hat{A}\}$ and $(\Delta\hat{A})^2_\text{mc}=\langle\hat{A}^2\rangle_\text{mc}-\langle\hat{A}\rangle_\text{mc}^2$ are the quantum expectation value and the quantum variance, respectively, in the microcanonical state
\begin{equation}
\hat{\rho}_\text{mc}
=\frac{1}{d}\sum_{j=1}^d\ket{E_j}\bra{E_j}.
\label{eqn:mc1}
\end{equation}
This implies that the probability of the expectation value $
\bra{\psi}\hat{A}\ket{\psi}$ in a sampled state $\ket{\psi}$ deviating from the microcanonical expectation value $\langle\hat{A}\rangle_\text{mc}$ is bounded by Chebyshev's inequality as
\begin{equation}
P\!\left(
|
\bra{\psi}\hat{A}\ket{\psi}
-\langle\hat{A}\rangle_\text{mc}
|^2>K(\Delta\hat{A})^2_\text{mc}
\right)<\frac{1}{K(d+1)}
\end{equation}
for any positive $K$.
Therefore, if the system is so large that the energy shell $\mathcal{H}_E$ contains many eigenstates $\ket{E_j}$, namely, if $d$ is large, a vast majority of the pure states $\ket{\psi}$ in the energy shell $\mathcal{H}_E$ exhibit expectation values close to the microcanonical expectation value:
\begin{equation}
\bra{\psi}\hat{A}\ket{\psi}
\simeq\langle\hat{A}\rangle_\text{mc}.
\end{equation}
This is the microcanonical typicality \cite{Sugita1b,*Sugita1,*Sugita1arXiv,Reimann1,Reimann2008-JSP,Sugiura1}.
Each typical pure state $\ket{\psi}$ in the energy shell $\mathcal{H}_E$ well describes the thermal equilibrium state of the system characterized by the energy $E$.

\section{NESS with a Small System}
\label{sec:NESS}
\begin{figure}[]
\begin{tabular}{l@{\qquad}c}
(a)&\\[-3.5truemm]
&\includegraphics[scale=0.35]{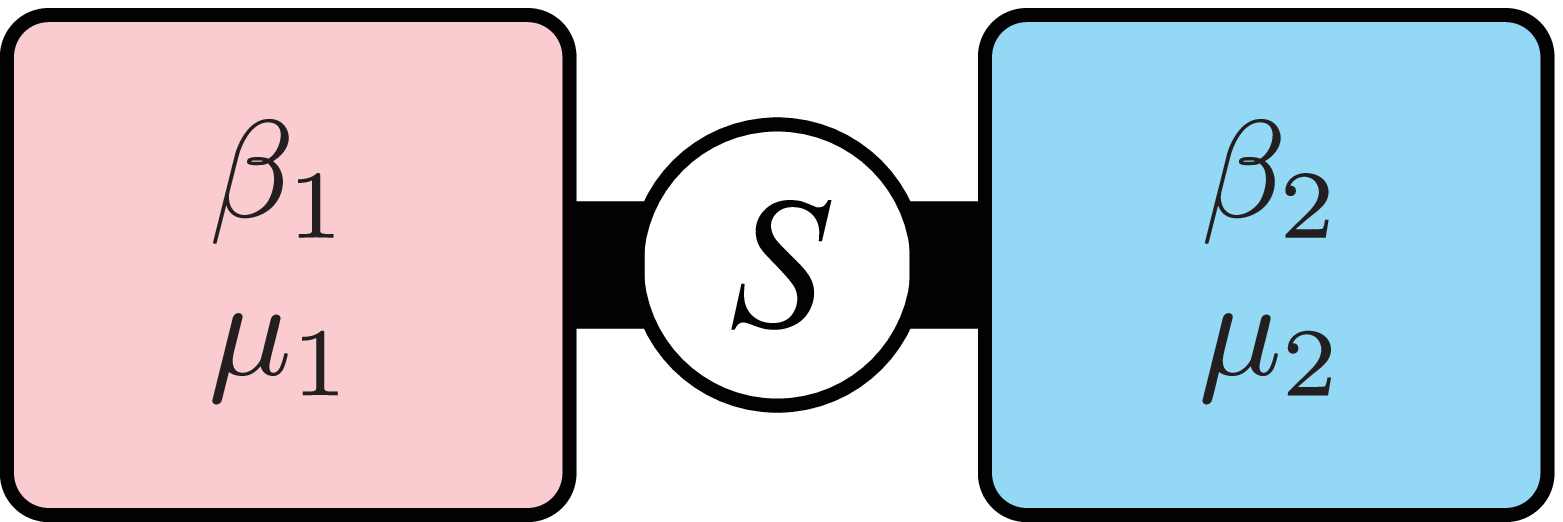}\\[8truemm]
(b)&\\[-3.5truemm]
&\includegraphics[scale=0.45]{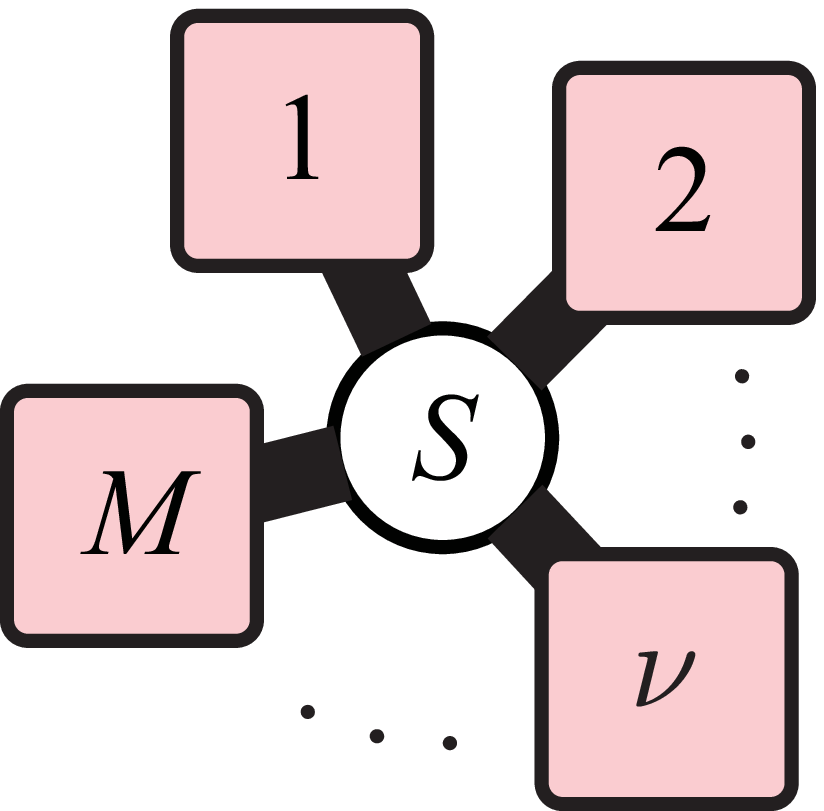}
\end{tabular}
\caption{(a) A small quantum system $S$ is coupled to two reservoirs at different inverse temperatures $\beta_\nu$ and different chemical potentials $\mu_\nu$ ($\nu=1,2$).  (b) A finite-dimensional quantum system $S$ is coupled to $M$ reservoirs at different inverse temperatures $\beta_\nu$ and different chemical potentials $\mu_\nu$ ($\nu=1,\ldots,M$).}
\label{fig:Setup}
\end{figure}

In this article, we are interested in NESS's, rather than equilibrium states.
In particular, we discuss NESS's realized in a quantum system involving a small (finite-dimensional) quantum system coupled to multiple (infinitely extended) large reservoirs \cite{ref:NESS-Ruelle,Tasaki3,ref:NESS-TasakiMatsui,ref:NESS-AschbacherPillet-JSP,ref:NESS-AschbacherJaksicPautratPillet,ref:NESS-Tasaki,Gaspard1,Saito1,ref:SaitoTasaki-ExtendedClausius}.
For instance, the simplest setup consists of a quantum dot placed between two reservoirs.
See Fig.\ \ref{fig:Setup}.
The reservoirs are characterized by different temperatures and different chemical potentials, and a steady current flows from reservoirs to reservoirs through the small system, say $S$, in a stationary state.

We first recall the standard approach to NESS's \cite{ref:NESS-Antal,ref:NESS-Ruelle,Tasaki3,ref:NESS-TasakiMatsui,ref:NESS-AschbacherPillet-JSP,ref:NESS-AschbacherJaksicPautratPillet,ref:NESS-Tasaki,Gaspard1,Saito1,ref:SaitoTasaki-ExtendedClausius}, based on statistical ensembles.
A natural way to construct a NESS is to let the system evolve and relax to a stationary state by itself.
For instance, suppose that the subsystems, i.e., small system $S$ and reservoirs $\nu\,(=1,\ldots,M)$, are initially disconnected from each other and the reservoirs are in local equilibrium.
The state of the total system is described by the density operator
\begin{equation}
\hat{\rho}_0=\hat{\rho}_S\otimes\left(\mathop{\bigotimes}_{\nu=1}^M\hat{\rho}_\text{gc}^{(\nu)}\right),
\label{eqn:Rho0Prod}
\end{equation}
where
\begin{equation}
\hat{\rho}_\text{gc}^{(\nu)}
=\frac{1}{\Xi_\nu}e^{-\beta_\nu(\hat{H}_\nu-\mu_\nu\hat{N}_\nu)}
\label{eqn:GC}
\end{equation}
is the grand canonical state of reservoir $\nu$, with $\beta_\nu$ being its inverse temperature, $\mu_\nu$ its chemical potential, $\hat{H}_\nu$ its free Hamiltonian, $\hat{N}_\nu$ the number of particles in reservoir $\nu$, and $\Xi_\nu$ its partition function, while the initial state $\hat{\rho}_S$ of system $S$ is arbitrary.
Then, at $t=0$ we let the total system start to evolve with the Hamiltonian 
\begin{equation}
\hat{H}=\hat{H}_0+\hat{V},\qquad
\hat{H}_0=\hat{H}_S+\sum_{\nu=1}^M\hat{H}_\nu,
\end{equation}
with $\hat{H}_S$ being the free Hamiltonian of system $S$, which admits only a finite number of different eigenstates, and $\hat{V}$ the interaction Hamiltonian describing the particle transfers between system $S$ and the reservoirs.
In the long-time limit $t\to\infty$, the total system would reach a stationary state,
\begin{equation}
e^{-i\hat{H}t}\hat{\rho}_0e^{i\hat{H}t}
\xrightarrow{t\to\infty}
\hat{\rho}_\text{NESS},
\label{eqn:Relax2NESS}
\end{equation}
which is a NESS\@.
In this stationary state, a current can flow steadily.
In particular, the expectation value of the current operator
\begin{equation}
\hat{J}_\nu=i[\hat{N}_\nu,\hat{H}],
\label{eqn:CurrentOp}
\end{equation}
describing the current flowing from reservoir $\nu$ into system $S$ may exhibit a nonzero value in the stationary state. 
This is the standard way to construct a NESS $\hat{\rho}_\text{NESS}$ \cite{ref:NESS-Antal,ref:NESS-Ruelle,Tasaki3,ref:NESS-TasakiMatsui,ref:NESS-AschbacherPillet-JSP,ref:NESS-AschbacherJaksicPautratPillet,ref:NESS-Tasaki,Gaspard1,Saito1,ref:SaitoTasaki-ExtendedClausius}, based on the grand canonical ensembles.

Note that we are considering infinitely extended reservoirs.
If the reservoirs are of finite size, the whole system might equilibrate in the long-time limit, becoming characterized by a single temperature and a single chemical potential.
We are not going to discuss such an equilibration process of a finite system in the present study.
We take the thermodynamical limit (the large-volume limit) before the long-time limit.
From a mathematical point of view the thermodynamical limit is important since the long-time limit requires a continuous spectrum.
The limits actually exist in various models, and there are a bunch of mathematically rigorous works on NESS's. See, e.g., \cite{ref:NESS-Ruelle,Tasaki3,ref:NESS-TasakiMatsui,ref:NESS-AschbacherPillet-JSP,ref:NESS-AschbacherJaksicPautratPillet,ref:NESS-Tasaki}.

Note also that the relaxation to a NESS is an irreversible process and the choice of the initial condition $\hat{\rho}_S$ for system $S$ becomes irrelevant, except for particular systems.
This irreversibility will be important for the discussion of the typicality for NESS\@.
In the following, we assume that the NESS is unique for a given setup and is independent of the initial state $\hat{\rho}_S$ of system $S$.
We will see later that it is actually the case for the model studied in this article.

\section{Construction of Pure NESS's}
\label{sec:Construction}
In the previous section, we have recalled the standard approach to describe NESS's in terms of statistical ensembles.
The objective of the present work is to show that there exist many \textit{pure} states $\ket{\phi}_\text{NESS}$ which can describe a NESS, i.e., there exist many pure states $\ket{\phi}_\text{NESS}$ exhibiting the expectation values of an observable $\hat{A}$ very close to that evaluated in the NESS $\hat{\rho}_\text{NESS}$,
\begin{equation}
\bras{\phi}{\text{NESS}}\hat{A}\ket{\phi}_\text{NESS}\simeq\Tr\{\hat{\rho}_\text{NESS}\hat{A}\}.
\label{eqn:PureMixedNESS}
\end{equation}
Such pure states $\ket{\phi}_\text{NESS}$ are regarded as \textit{pure NESS's}.
Moreover, they are just \textit{typical} states of a large Hilbert space $\mathcal{H}_\text{NESS}$ identified below: a pure state randomly sampled from $\mathcal{H}_\text{NESS}$ almost surely represents the NESS as (\ref{eqn:PureMixedNESS}).

In Ref.\ \cite{ref:TypicalNESS}, we have provided the construction of typical pure NESS's $\ket{\phi}_\text{NESS}$ in the absence of the small system $S$ between the reservoirs.
The idea is to sample a typical pure state $\ket{\phi}$ to represent the initial thermal equilibrium states $\mathop{\bigotimes}_\nu\hat{\rho}_\text{gc}^{(\nu)}$ of the multiple reservoirs $\nu\,(=1,\ldots,M)$, 
and let it evolve by the Hamiltonian $\hat{H}$ of the system to a stationary state in the long-time limit $t\to\infty$ to get a typical pure NESS $\ket{\phi}_\text{NESS}$.
Since the reservoirs are initially in local equilibrium, each of which is characterized by the energy $E_\nu$ corresponding to its inverse temperature $\beta_\nu$ and its chemical potential $\mu_\nu$, the relevant Hilbert space from which the initial pure state $\ket{\phi}$ is to be sampled is $\mathop{\bigotimes}_\nu\mathcal{H}_{E_\nu}^{(\nu)}$, where $\mathcal{H}_{E_\nu}^{(\nu)}$ is the Hilbert space of reservoir $\nu$ spanned by the eigenstates of $\hat{H}_\nu$ belonging to the energies within the energy shell $[E_\nu,E_\nu+\Delta E_\nu]$, with the number of particles fixed at $N_\nu$.
Now, in the presence of the small system $S$ between the reservoirs, what is the relevant Hilbert space from which the initial state of $S$ is to be sampled?
We will see that the answer is the whole Hilbert space $\mathcal{H}_S$ of $S$: we do not need to restrict ourselves to some energy shell of $S$.

The recipe for constructing the pure NESS's $\ket{\phi}_\text{NESS}$ in the presence of the small system $S$ is the following.
We pick a pure state $\ket{\phi}$ randomly from the Hilbert space $\mathcal{H}_{S,\{E_\nu\}}$, 
\begin{equation}
\ket{\phi}\in\mathcal{H}_{S,\{E_\nu\}}=\mathcal{H}_S\otimes\left(\mathop{\bigotimes}_{\nu=1}^M\mathcal{H}_{E_\nu}^{(\nu)}\right),
\label{eqn:Hprod}
\end{equation}
according to the Haar measure.
We then ``scatter'' it,
\begin{equation}
\ket{\phi}_\text{NESS}
=\hat{W}\ket{\phi}
\label{eqn:PureNESSConst}
\end{equation}
by the M\o ller wave operator \cite{Thirring1,ref:ScatteringTaylor}
\begin{equation}
\hat{W}=\lim_{t\to\infty}e^{-i\hat{H}t}e^{i\hat{H}_0t},
\end{equation}
to get a stationary state.
We are going to prove that the pure states $\ket{\phi}_\text{NESS}$ constructed in this way almost surely give the expectation values of an observable $\hat{A}$ close to that evaluated in $\hat{\rho}_\text{NESS}$ as in (\ref{eqn:PureMixedNESS}), and are regarded as pure NESS's.
Such pure NESS's $\ket{\phi}_\text{NESS}$ are typical states of the Hilbert space $\mathcal{H}_\text{NESS}$, which is isometric to $\mathcal{H}_{S,\{E_\nu\}}$ through the wave operator $\hat{W}$ \cite{Thirring1,ref:ScatteringTaylor}.
A vast majority of the pure states in $\mathcal{H}_\text{NESS}$ well describe the NESS\@.

\section{Typicality of Pure NESS's}
\label{sec:TypicalNESS}
Let us prove that the pure states $\ket{\phi}_\text{NESS}$ constructed by (\ref{eqn:PureNESSConst}) actually represent a NESS\@.

First, we sample a pure state $\ket{\phi}$ from $\mathcal{H}_{S,\{E_\nu\}}$ to represent the initial state $\rho_0$ in (\ref{eqn:Rho0Prod}),
\begin{multline}
\ket{\phi}
=\sum_{i=1}^{d_S}\sum_{j_1=1}^{d_1}\cdots\sum_{j_M=1}^{d_M}
c_{ij_1\ldots j_M}
\ket{E_i^{(S)}}
\otimes\ket{E_{j_1}^{(1)}}
\\
{}\otimes\cdots\otimes\ket{E_{j_M}^{(M)}},
\label{eqn:PhiEnt}
\end{multline}
where $\ket{E_i^{(S)}}\in\mathcal{H}_S$ are orthonormal basis states of system $S$ and $\ket{E_{j_\nu}^{(\nu)}}\in\mathcal{H}_{E_\nu}^{(\nu)}$ are the energy eigenstates of $\hat{H}_\nu$ belonging to the energies $E_{j_\nu}^{(\nu)}$ within the energy shell $[E_\nu,E_\nu+\Delta E_\nu]$, while $d_S=\dim\mathcal{H}_S$ and $d_\nu=\dim\mathcal{H}_{E_\nu}^{(\nu)}$ are the dimensions of the Hilbert spaces.
The dimensions $d_\nu$ of the reservoirs are supposed to be finite for the moment, but the thermodynamical limit will be taken later at certain point.

Note that the pure state $\ket{\phi}$ of the form (\ref{eqn:PhiEnt}) is not a product state, but is highly entangled in general.
A naive way to represent the initial state $\hat{\rho}_0$ in (\ref{eqn:Rho0Prod}) by a pure state would be to represent each individual thermal equilibrium state $\hat{\rho}_\text{gc}^{(\nu)}$ in the product state (\ref{eqn:Rho0Prod}) separately by a typical pure state on its Hilbert space $\mathcal{H}_{E_\nu}^{(\nu)}$, on the basis of the knowledge on the typicality for equilibrium systems recapitulated in Sec.\ \ref{sec:TypicalEq}\@.
The state $\ket{\phi}$ of the total system constructed in this way is a product state, with no entanglement among system $S$ and the reservoirs.
The set of such pure product states occupy only a small portion of the Hilbert space $\mathcal{H}_{S,\{E_\nu\}}$ in (\ref{eqn:Hprod}).
By (\ref{eqn:PhiEnt}), we explore a much larger Hilbert space, i.e., the whole Hilbert space $\mathcal{H}_{S,\{E_\nu\}}$.

It is not trivial whether the highly entangled states $\ket{\phi}$ of the form (\ref{eqn:PhiEnt}) represents the product state $\hat{\rho}_0$ in (\ref{eqn:Rho0Prod}).
Let us prove it.
We sample the pure states $\ket{\phi}$ uniformly from $\mathcal{H}_{S,\{E_\nu\}}$ according to the Haar measure [cf.\ (\ref{eqn:HaarEq})]
\begin{multline}
d\mu(\ket{\phi})
\propto
\delta\!\left(
\sum_{i=1}^{d_S}\sum_{j_1=1}^{d_1}\cdots\sum_{j_M=1}^{d_M}
|c_{ij_1\ldots j_M}|^2-1
\right)
\\
{}\times
\prod_{i=1}^{d_S}\prod_{j_1=1}^{d_1}\cdots\prod_{j_M=1}^{d_M}d^2c_{ij_1\ldots j_M},
\end{multline}
which yields [cf.\ (\ref{eqn:Cmoments})]
\begin{subequations}
\begin{equation}
\overline{|c_{ij_1\ldots j_M}|^2}=\frac{1}{D},
\end{equation}
\begin{multline}
\overline{|c_{ij_1\ldots j_M}|^2|c_{i'j_1'\ldots j_M'}|^2}
\\
=\frac{1}{D(D+1)}
(1+\delta_{ii'}\delta_{j_1j_1'}\cdots\delta_{j_Mj_M'}),
\end{multline}
\end{subequations}
with the others up to the fourth moments of the coefficients $c_{ij_1\ldots j_M}$ vanishing, where 
$D=d_Sd_1\cdots d_M$ is the dimension of the total Hilbert space $\mathcal{H}_{S,\{E_\nu\}}$.
Then, we get [cf.\ (\ref{eqn:AveVarEq}) and \cite{ref:TypicalNESS}]
\begin{equation}
\overline{\bra{\phi}\hat{A}\ket{\phi}}
=\langle\hat{A}\rangle_\text{mc},\quad
\Var[\bra{\phi}\hat{A}\ket{\phi}]
\le\frac{(\Delta\hat{A})^2_\text{mc}}{D+1},
\end{equation}
where $\langle\hat{A}\rangle_\text{mc}$ and $(\Delta\hat{A})^2_\text{mc}$ are evaluated in the microcanonical state given [instead of (\ref{eqn:mc1})] by
\begin{equation}
\hat{\rho}_\text{mc}=\hat{\rho}_\text{mc}^{(S)}\otimes\left(\mathop{\bigotimes}_{\nu=1}^M\hat{\rho}_\text{mc}^{(\nu)}\right)
\label{eqn:RhoMCProd}
\end{equation}
with
\begin{multline}
\hat{\rho}_\text{mc}^{(S)}=\frac{1}{d_S}\openone_S,\quad
\hat{\rho}_\text{mc}^{(\nu)}=\frac{1}{d_\nu}\sum_{j_\nu=1}^{d_\nu}\ket{E_{j_\nu}^{(\nu)}}\bra{E_{j_\nu}^{(\nu)}}
\\
(\nu=1,\ldots,M).
\label{eqn:MCstate}
\end{multline}
This result implies that for large reservoirs (i.e., for large $D$) the pure states $\ket{\phi}$ sampled from $\mathcal{H}_{S,\{E_\nu\}}$ typically yield
\begin{equation}
\bra{\phi}\hat{A}\ket{\phi}
\simeq\langle\hat{A}\rangle_\text{mc}
\label{eqn:TypicalityMC}
\end{equation}
with vanishingly small errors.
Through the equivalence between the microcanonical ensembles $\hat{\rho}_\text{mc}^{(\nu)}$ and the grand canonical ensembles $\hat{\rho}_\text{gc}^{(\nu)}$ for large reservoirs, we have
\begin{equation}
\bra{\phi}\hat{A}\ket{\phi}
\simeq\langle\hat{A}\rangle_\text{gc},
\label{eqn:TypicalityPhi}
\end{equation}
where $\langle\hat{A}\rangle_\text{gc}$ is evaluated in the state
\begin{equation}
\hat{\rho}_\text{gc}=\hat{\rho}_\text{mc}^{(S)}\otimes\left(\mathop{\bigotimes}_{\nu=1}^M\hat{\rho}_\text{gc}^{(\nu)}\right).
\label{eqn:GCmix}
\end{equation}
This proves that the typical pure states $\ket{\phi}$ in $\mathcal{H}_{S,\{E_\nu\}}$ well represent the state $\hat{\rho}_\text{gc}$ defined in (\ref{eqn:GCmix}), and hence the local equilibrium states $\mathop{\bigotimes}_\nu\hat{\rho}_\text{gc}^{(\nu)}$ of the reservoirs in the initial product state $\hat{\rho}_0$ in (\ref{eqn:Rho0Prod}).

On the other hand, the initial state $\hat{\rho}_S$ of the small system $S$ in $\hat{\rho}_0$ is not reproduced in $\hat{\rho}_\text{gc}$ in (\ref{eqn:GCmix}).
This is, however, not a problem in constructing the typical pure NESS's $\ket{\phi}_\text{NESS}$, \textit{provided the system admits a unique NESS independent of the initial condition $\hat{\rho}_S$ for the small system $S$}.
Indeed, the state $\ket{\phi}_\text{NESS}$ constructed by (\ref{eqn:PureNESSConst}) with a typical pure state $\ket{\phi}$ of $\mathcal{H}_{S,\{E_\nu\}}$ yields
\begin{align}
&\bras{\phi}{\text{NESS}}\hat{A}\ket{\phi}_\text{NESS}
\nonumber\\
&\qquad
=\bra{\phi}\hat{W}^\dag\hat{A}\hat{W}\ket{\phi}
\nonumber\\
&\qquad
\simeq\langle\hat{W}^\dag\hat{A}\hat{W}\rangle_\text{gc}
\nonumber\\
&\qquad
=\lim_{t\to\infty}\Tr\{
e^{-i\hat{H}t}
\hat{\rho}_\text{gc}
e^{i\hat{H}t}
\hat{A}
\}
\nonumber\\
&\qquad
=\Tr\{
\hat{\rho}_\text{NESS}
\hat{A}
\},
\label{eqn:TypicalityNESS}
\end{align}
where we have used the typicality of $\ket{\phi}$ in (\ref{eqn:TypicalityPhi}), the fact that $\hat{\rho}_\text{gc}$ defined in (\ref{eqn:GCmix}) is stationary under the action of $\hat{H}_0$, and the relaxation to the unique NESS $\hat{\rho}_\text{NESS}$ in (\ref{eqn:Relax2NESS}).
Even if the state $\hat{\rho}_\text{mc}^{(S)}$ of the small system $S$ in $\hat{\rho}_\text{gc}$ is different from $\hat{\rho}_S$ in $\hat{\rho}_0$, the two initial conditions $\hat{\rho}_\text{gc}$ and $\hat{\rho}_0$ yield the same NESS $\hat{\rho}_\text{NESS}$ in the long-time limit, since we are assuming that the NESS is unique and independent of the initial condition for $S$.

Equation (\ref{eqn:TypicalityNESS}) holds for any typical $\ket{\phi}$ in $\mathcal{H}_{S,\{E_\nu\}}$, and shows the typicality of the pure NESS's $\ket{\phi}_\text{NESS}$.
Note that $\ket{\phi}_\text{NESS}$ is isometric to $\ket{\phi}$, connected by the M\o ller wave operator $\hat{W}$ \cite{Thirring1,ref:ScatteringTaylor}.
Therefore, the pure NESS's $\ket{\phi}_\text{NESS}$ are typical states in the Hilbert space $\mathcal{H}_\text{NESS}$ which is isometric to $\mathcal{H}_{S,\{E_\nu\}}$ through $\hat{W}$.

As stressed in Sec.\ \ref{sec:NESS},  we are interested in infinitely large reservoirs.
But we actually start by sampling the typical pure state $\ket{\phi}$ from the Hilbert space $\mathcal{H}_{S,\{E_\nu\}}$ of finite dimension $D=d_Sd_1\cdots d_M$.
For large $D$, the typicality (\ref{eqn:TypicalityPhi}) holds, and the expectation value in $\ket{\phi}$ is replaced by that in $\hat{\rho}_\text{gc}$ defined in (\ref{eqn:GCmix}), under the equivalence between the microcanonical and grand canonical ensembles.
We then take the thermodynamical limit to assure the existence of the long-time limit to get the NESS $\hat{\rho}_\text{NESS}$ in (\ref{eqn:TypicalityNESS}).
The thermodynamical limit should be taken before the long-time limit $t\to\infty$.

For such infinitely extended systems, interesting observables $\hat{A}$ [e.g., the current operators $\hat{J}_\nu$ in (\ref{eqn:CurrentOp})] would be unbounded operators.
It is, however, not a problem for the typicality.
It is clear from the proofs of the typicality in Sec.\ \ref{sec:TypicalEq} and in the present section that what is crucial for the typicality is not the boundedness of the observable $\hat{A}$ but the finiteness of its expectation value $\langle\hat{A}\rangle_\text{mc}$ and the variance $(\Delta\hat{A})^2_\text{mc}$ in the relevant microcanonical state: even if the observable $\hat{A}$ is an unbounded operator, the typicality holds as long as the microcanonical expectation value and the microcanonical variance of $\hat{A}$ are finite.
It is the case for thermodynamically relevant quantities (intensive quantities, or extensive quantities per volume).

Finally, it would be practically easier to work in the Heisenberg picture than in the Schr\"odinger picture to analyze large quantum many-body systems.
When calculating the expectation value $\bras{\phi}{\text{NESS}}\hat{A}\ket{\phi}_\text{NESS}$ of an observable $\hat{A}$ in a typical pure NESS $\ket{\phi}_\text{NESS}$, we would scatter the observable $\hat{A}$ as $\hat{W}^\dag\hat{A}\hat{W}$ instead of scattering a typical pure state $\ket{\phi}$, and evaluate its expectation value $\bra{\phi}\hat{W}^\dag\hat{A}\hat{W}\ket{\phi}$ in the initial typical pure state $\ket{\phi}$.
The typicality of pure NESS's $\ket{\phi}_\text{NESS}$ is then reduced to the typicality of $\ket{\phi}$ for equilibrium systems.
In the next section, we will study a model in this way.

\section{Model}
\label{sec:Model}
As stressed in the previous section, the irreversible relaxation to a unique NESS independent of the initial state $\hat{\rho}_S$ of the small system $S$ is important for the typicality of the pure NESS's $\ket{\phi}_\text{NESS}$ constructed in Sec.\ \ref{sec:Construction}\@.
Let us here look at an example.
As we will see below, perturbative treatment is not useful for the discussion of the NESS\@.
Let us hence look at an exactly solvable model.

We consider a quantum dot $S$ coupled to $M$ fermionic reservoirs.
The Hamiltonian reads
\begin{subequations}
\label{eqn:Model}
\begin{equation}
\hat{H}=\hat{H}_0+\hat{V},\qquad
\hat{H}_0=\hat{H}_S+\sum_{\nu=1}^M\hat{H}_\nu
\end{equation}
with
\begin{gather}
\hat{H}_S
=\Omega\hat{a}^\dag\hat{a},\quad
\hat{H}_\nu=\sum_{\nu=1}^M\int d^3\bm{k}\,\omega_{k\nu}\hat{b}_{\bm{k}\nu}^\dag\hat{b}_{\bm{k}\nu},
\label{eqn:H0}
\displaybreak[0]\\
\hat{V}
=\lambda\sum_{\nu=1}^M\int d^3\bm{k}\,(u_{\bm{k}\nu}^*\hat{a}^\dag\hat{b}_{\bm{k}\nu}
+u_{\bm{k}\nu}\hat{b}_{\bm{k}\nu}^\dag\hat{a}),
\end{gather}
\end{subequations}
where $\hat{a}$ and $\hat{b}_{\bm{k}\nu}$ are fermionic operators satisfying the canonical anticommunication relations $\{\hat{a},\hat{a}^\dag\}=1$, $\{\hat{b}_{\bm{k}\nu},\hat{b}_{\bm{k}'\nu'}^\dag\}=\delta_{\nu\nu'}\delta^3(\bm{k}-\bm{k}')$, $\{\hat{a},\hat{a}\}=\{\hat{b}_{\bm{k}\nu},\hat{b}_{\bm{k}'\nu'}\}=\{\hat{a},\hat{b}_{\bm{k}\nu}\}=\{\hat{a},\hat{b}_{\bm{k}\nu}^\dag\}=0$, and we assume that $\Omega,\omega_{k\nu}>0$.
We can also think of a bosonic system, with the same Hamiltonian as (\ref{eqn:Model}) but with bosonic operators $\hat{a}$ and $\hat{b}_{\bm{k}\nu}$ satisfying the canonical communication relations $[\hat{a},\hat{a}^\dag]=1$, $[\hat{b}_{\bm{k}\nu},\hat{b}_{\bm{k}'\nu'}^\dag]=\delta_{\nu\nu'}\delta^3(\bm{k}-\bm{k}')$, $[\hat{a},\hat{a}]=[\hat{b}_{\bm{k}\nu},\hat{b}_{\bm{k}'\nu'}]=[\hat{a},\hat{b}_{\bm{k}\nu}]=[\hat{a},\hat{b}_{\bm{k}\nu}^\dag]=0$.
All the following formulas are valid in both fermionic and bosonic cases apart from a few signs (upper/lower signs are for fermionic/bosonic case in the following formulas).

It should be noted however that in the previous sections system $S$ is assumed to be finite-dimensional, admitting only a finite number of energy levels.
It is actually the case in the fermionic case, while it is not for the bosonic oscillator $\hat{a}$.
Nonetheless, it is not crucial.
For the bosonic oscillator, we just have to restrict $\mathcal{H}_{S,\{E_\nu\}}$ in (\ref{eqn:Hprod}), from which a typical pure state $\ket{\phi}$ is sampled to construct a typical pure NESS by (\ref{eqn:PureNESSConst}): we replace $\mathcal{H}_S$ in $\mathcal{H}_{S,\{E_\nu\}}$ with a finite-dimensional subspace $\tilde{\mathcal{H}}_S$ of $\mathcal{H}_S$, e.g., by introducing an energy cutoff in the initial state of $S$.
Due to the irreversibility of the dynamics, such details in the initial state of $S$ is irrelevant to the construction of typical pure NESS's.
Note that by restricting the Hilbert space as $\tilde{\mathcal{H}}_S$ we do not mean to truncate the Hamiltonian $H_S$ of $S$: we just restrict the Hilbert space from which the initial state $\ket{\phi}$ is sampled, while the oscillator $S$ can evolve over the whole Hilbert space $\mathcal{H}_S$ with no restriction.

\subsection{Heisenberg Picture}
\label{sec:HeisenbergOps}
The model (\ref{eqn:Model}) is solvable exactly.
Indeed, the Heisenberg equations of motion for the Heisenberg operators $\hat{a}(t)=e^{i\hat{H}t}\hat{a}e^{-i\hat{H}t}$ and $\hat{b}_{\bm{k}\nu}(t)=e^{i\hat{H}t}\hat{b}_{\bm{k}\nu}e^{-i\hat{H}t}$,
\begin{subequations}
\label{eqn:HeisenbergEq}
\begin{gather}
\frac{d}{dt}\hat{a}(t)
=-i\Omega\hat{a}(t)-i\lambda\sum_{\nu=1}^M\int d^3\bm{k}\,u_{\bm{k}\nu}^*\hat{b}_{\bm{k}\nu}(t),
\displaybreak[0]\\
\frac{d}{dt}\hat{b}_{\bm{k}\nu}(t)
=-i\omega_{k\nu}\hat{b}_{\bm{k}\nu}(t)-i\lambda u_{\bm{k}\nu}\hat{a}(t),
\end{gather}
\end{subequations}
are solvable exactly, yielding
\begin{subequations}
\label{eqn:SolHeisenbergOps}
\begin{gather}
\hat{a}(t)
=G(t)\hat{a}-i\lambda\int_0^tdt'\,G(t-t')\hat{B}(t'),
\displaybreak[0]\\
\hat{b}_{\bm{k}\nu}(t)
=e^{-i\omega_{k\nu}t}\hat{b}_{\bm{k}\nu}-i\lambda\int_0^tdt'\,e^{-i\omega_{k\nu}(t-t')}u_{\bm{k}\nu}\hat{a}(t'),
\end{gather}
\end{subequations}
where
\begin{equation}
\hat{B}(t)
=\sum_{\nu=1}^M\int d^3\bm{k}\,u_{\bm{k}\nu}^*e^{-i\omega_{k\nu}t}\hat{b}_{\bm{k}\nu},
\end{equation}
and
\begin{subequations}
\label{eqn:Green}
\begin{gather}
G(t)=\int_0^{\infty}\frac{d\omega}{2\pi}\tilde{G}(\omega)e^{-i\omega t},
\displaybreak[0]\\
\tilde{G}(\omega)=\frac{\lambda^2\Gamma(\omega)}{[\omega-\Omega-\lambda^2\Delta(\omega)]^2+[\lambda^2\Gamma(\omega)/2]^2},
\label{eqn:GFourier}
\end{gather}
\end{subequations}
with
\begin{gather}
\Gamma(\omega)
=2\pi\sum_{\nu=1}^M\int d^3\bm{k}\,
|u_{\bm{k}\nu}|^2
\delta(\omega_{k\nu}-\omega),
\label{eqn:Gamma}
\displaybreak[0]\\
\Delta(\omega)
=\pv\int_0^\infty\frac{d\omega'}{2\pi}\frac{\Gamma(\omega')}{\omega-\omega'}.
\end{gather}
In obtaining the Fourier representation (\ref{eqn:Green}), we have assumed that the total Hamiltonian $H$ does not admit a bound state.
It is actually the case if the coupling $\lambda$ is not too strong, below a threshold value \cite{ref:Miyamoto-BoundStatesPRA2005}.

\subsection{NESS}
\label{sec:NESSmodel}
Let us look at a NESS in the present model, on the basis of the standard ensemble approach recapitulated in Sec.\ \ref{sec:NESS}\@.
For the initial state $\rho_0$ given in (\ref{eqn:Rho0Prod}), with the local equilibrium states of the reservoirs described by the grand canonical ensembles $\hat{\rho}_\text{gc}^{(\nu)}$ given in (\ref{eqn:GC}), the characteristic function of the total system is computable exactly.
By noting 
\begin{gather}
\langle
\hat{b}_{\bm{k}\nu}
\rangle_0
=0,\ \ %
\langle
\hat{b}_{\bm{k}\nu}^\dag
\hat{b}_{\bm{k}'\nu'}
\rangle_0
=f_\nu(\omega_{k\nu})\delta_{\nu\nu'}\delta^3(\bm{k}-\bm{k}'),\ \ %
\text{etc.}
\end{gather}
in the initial state $\rho_0$, with the Fermi/Bose distribution function
\begin{equation}
f_\nu(\omega)
=\frac{1}{e^{\beta_\nu(\omega-\mu_\nu)}\pm1},
\end{equation}
we get \cite{ref:FactorizeModel}
\begin{widetext}
\begin{align}
\chi_t[\xi,\xi^*,\eta,\eta^*]
={}&\langle
e^{\hat{a}^\dag\xi-\xi^*\hat{a}+\sum_\nu\int d^3\bm{k}\,(\hat{b}_{\bm{k}\nu}^\dag\eta_{\bm{k}\nu}-\eta_{\bm{k}\nu}^*\hat{b}_{\bm{k}\nu})}\rangle_t
\nonumber\displaybreak[0]\\
={}&\langle
e^{\hat{a}^\dag(t)\xi-\xi^*\hat{a}(t)+\sum_\nu\int d^3\bm{k}\,[\hat{b}_{\bm{k}\nu}^\dag(t)\eta_{\bm{k}\nu}-\eta_{\bm{k}\nu}^*\hat{b}_{\bm{k}\nu}(t)]}\rangle_0
\nonumber\displaybreak[0]\\
={}&
\chi_S\bm{(}\xi(t),\xi^*(t)\bm{)}
\exp\!\left(
-\frac{1}{2}
\lambda^2\int_0^tdt_1\int_0^tdt_2\,
\xi^*G(t_1)
K^\beta(t_2-t_1)
G^*(t_2)\xi
\right)
\nonumber\displaybreak[0]\\
&
{}\times
\exp\biggl(
-\frac{1}{2}
K_{\eta\eta}^\beta(0)
+\lambda^2\Re\int_0^tdt'\,
K_\eta^\beta(t')
(G^*\circ K_\eta^*)(t')
\nonumber\displaybreak[0]\\
&
\qquad\qquad\qquad\qquad{}
-\frac{1}{2}
\lambda^4\int_0^tdt_1\int_0^tdt_2\,
(K_\eta\circ G)(t_1)
K^\beta(t_2-t_1)
(G^*\circ K_\eta^*)(t_2)
\biggr)
\nonumber\displaybreak[0]\\
&
{}\times
\exp\!\left(
\lambda\Im\int_0^tdt'\,
K_\eta^\beta(t')
G^*(t')\xi
-\lambda^3\Im\int_0^tdt_1\int_0^tdt_2\,
(K_\eta\circ G)(t_1)
K^\beta(t_2-t_1)
G^*(t_2)\xi
\right),
\label{eqn:CharFunc}
\end{align}
\end{widetext}
where $\xi$, $\xi^*$, $\eta_{\bm{k}\nu}$, $\eta_{\bm{k}\nu}^*$ are Grassmann variables in the fermionic case while they are just normal variables in the bosonic case,
\begin{equation}
\chi_S(\xi,\xi^*)
=\langle
e^{\hat{a}^\dag\xi-\xi^*\hat{a}}
\rangle_0
\end{equation}
is the characteristic function for the initial state $\rho_S$ of system $S$, and
\begin{equation}
\xi^*(t)
=\xi^*G(t)-i\lambda (K_\eta\circ G)(t),
\label{eqn:XiT}
\end{equation}
with $
(F\circ G)(t)=\int_0^tdt'\,F(t-t')G(t')
$ denoting convolution and the kernel functions
\begin{gather}
K_\flat^\sharp(t)
=\int_0^\infty\frac{d\omega}{2\pi}
\Gamma_\flat^\sharp(\omega)
e^{-i\omega t}
\quad
\begin{cases}
\smallskip
\sharp=\text{null or}\ \beta,\\
\flat=\text{null or}\ \eta\ \text{or}\ \eta\eta,
\end{cases}
\label{eqn:KernelFuncs}
\end{gather}
being given in terms of the spectral functions
\begin{gather}
\Gamma_\flat(t)
=\sum_{\nu=1}^M\Gamma_\flat^{(\nu)}(\omega),\quad
\Gamma_\flat^\beta(t)
=\sum_{\nu=1}^M[1\mp f_\nu(\omega)]\Gamma_\flat^{(\nu)}(\omega)
\label{eqn:SpectralFuncs}
\end{gather}
with 
\begin{subequations}
\begin{align}
&\Gamma^{(\nu)}(\omega)
=2\pi\int d^3\bm{k}\,
|u_{\bm{k}\nu}|^2
\delta(\omega_{k\nu}-\omega),
\label{eqn:GammaNu}
\displaybreak[0]\\
&\Gamma_\eta^{(\nu)}(\omega)
=2\pi\int d^3\bm{k}\,
\eta_{\bm{k}\nu}^*
u_{\bm{k}\nu}
\delta(\omega_{k\nu}-\omega),
\displaybreak[0]\\
&\Gamma_{\eta\eta}^{(\nu)}(\omega)
=2\pi\int d^3\bm{k}\,
\eta_{\bm{k}\nu}^*
\eta_{\bm{k}\nu}
\delta(\omega_{k\nu}-\omega).
\end{align}
\end{subequations}
The characteristic function $\chi_t[\xi,\xi^*,\eta,\eta^*]$ in (\ref{eqn:CharFunc}) is exact and characterizes the state of the total system at any time $t$ starting from the initial state $\rho_0$ given in (\ref{eqn:Rho0Prod}).

Now, notice that the Green function $G(t)$ in (\ref{eqn:Green}) and the kernel function $K_\eta(t)$ in (\ref{eqn:KernelFuncs}) are both decaying functions of time $t$ \cite{ref:Paolo-HydrogenPLA1998,ref:Paolo-VanHovePHA1999}, according to the Riemann-Lebesgue lemma.
Their convolution $(K_\eta\circ G)(t)$ also decays, and therefore, $\xi(t)$ in (\ref{eqn:XiT}) decays to zero $\xi(t)\to0$ in the long-time limit $t\to\infty$.
As a consequence, since $\chi_S\bm{(}\xi(t),\xi^*(t)\bm{)}\to\chi_S(0,0)=1$, the characteristic function $\chi_t[\xi,\xi^*,\eta,\eta^*]$ in (\ref{eqn:CharFunc}) becomes independent of the initial state $\rho_S$ of system $S$ 
and approaches
\begin{widetext}
\begin{align}
\chi_\text{NESS}[\xi,\xi^*,\eta,\eta^*]
={}&
\exp\!\left(
-\frac{1}{2}
\xi^*
\xi
\int_0^\infty\frac{d\omega}{2\pi}
\tilde{G}(\omega)
\frac{\Gamma^\beta(\omega)}{\Gamma(\omega)}
\right)
\nonumber\\
&{}\times
\exp\Biggl(
-\frac{1}{2}
\int_0^\infty\frac{d\omega}{2\pi}
\Gamma_{\eta\eta}^\beta(\omega)
+\lambda^2
\Re
\int_0^\infty\frac{d\omega}{2\pi}
\tilde{G}(\omega)
\int_0^\infty\frac{d\omega'}{2\pi}
\frac{
\Gamma_\eta^\beta(\omega')
}{\omega-\omega'+i0^+}
\int_0^\infty\frac{d\omega''}{2\pi}
\frac{
\Gamma_\eta^*(\omega'')
}{\omega'-\omega''-i0^+}
\nonumber\displaybreak[0]\\
&\qquad\qquad\qquad\qquad\quad\ \ 
{}-\frac{1}{2}
\lambda^2\int_0^\infty\frac{d\omega}{2\pi}
\tilde{G}(\omega)
\frac{\Gamma^\beta(\omega)}{\Gamma(\omega)}
\int_0^\infty\frac{d\omega'}{2\pi}
\frac{\Gamma_\eta(\omega')}{\omega-\omega'+i0^+}
\int_0^\infty\frac{d\omega''}{2\pi}
\frac{\Gamma_\eta^*(\omega'')}{\omega-\omega''-i0^+}
\Biggr)
\nonumber\displaybreak[0]\\
&{}\times
\exp\!\left[
\lambda
\Re
\int_0^\infty\frac{d\omega}{2\pi}
\tilde{G}(\omega)
\int_0^\infty\frac{d\omega'}{2\pi}
\left(
\Gamma_\eta^\beta(\omega')
-\Gamma_\eta(\omega')
\frac{\Gamma^\beta(\omega)}{\Gamma(\omega)}
\right)
\frac{
1
}{\omega-\omega'+i0^+}
\xi\right].
\label{eqn:CharFuncNESS}
\end{align}
\end{widetext}
This is the exact characteristic function of the NESS\@.
The first exponential factor characterizes the state of system $S$, while the second characterizes the state of the reservoirs. 
The third one describes the correlations between system $S$ and the reservoirs, and is relevant to the current flowing steadily between $S$ and the reservoirs.
For the present model, the number of particles in reservoir $\nu$ is given by
\begin{equation}
\hat{N}_\nu=\int d^3\bm{k}\,\hat{b}_{\bm{k}\nu}^\dag\hat{b}_{\bm{k}\nu},
\end{equation}
and the current operator $\hat{J}_\nu$ in (\ref{eqn:CurrentOp}) describing the current from reservoir $\nu$ into system $S$ reads
\begin{equation}
\hat{J}_\nu
=-i\lambda\int d^3\bm{k}\,(u_{\bm{k}\nu}^*\hat{a}^\dag\hat{b}_{\bm{k}\nu}-u_{\bm{k}\nu}\hat{b}_{\bm{k}\nu}^\dag\hat{a}),
\label{eqn:CurrentOpModel}
\end{equation}
whose expectation value in the NESS is generated from the characteristic function (\ref{eqn:CharFuncNESS}) to be
\begin{align}
&\langle\hat{J}_\nu\rangle_\text{NESS}
\nonumber\\
&\ =
\lambda^2
\sum_{\nu'=1}^M
\int_0^\infty\frac{d\omega}{2\pi}
\tilde{G}(\omega)
\frac{
\Gamma^{(\nu)}(\omega)
\Gamma^{(\nu')}(\omega)
}{\Gamma(\omega)}
[
f_\nu(\omega)
-f_{\nu'}(\omega)
].
\label{eqn:NESSCurrent}
\end{align}
In this way, the present model admits the NESS, which is independent of the initial state $\hat{\rho}_S$ of system $S$, and the steady current flows through $S$.

It is worth noting here that the naive perturbative treatment is not useful for the analysis, unlike the case studied in Ref.\ \cite{ref:TypicalNESS}, in which the small system $S$ is absent between the reservoirs.
This is clear from the fact that the spectral function $\Gamma(\omega)$, which is quadratic in the coupling functions $u_{\bm{k}\nu}$, is found in denominators in the exponents in (\ref{eqn:CharFuncNESS}).
To properly capture the weak-coupling regime, van Hove's limit, $\lambda\to0$ keeping $\tau=\lambda^2t$ finite, is helpful.
In this limit, we have \cite{ref:Paolo-VanHovePHA1999}
\begin{equation}
\bar{G}(\tau/\lambda^2)
=G(\tau/\lambda^2)e^{i\Omega\tau/\lambda^2}
\to e^{-[\Gamma(\Omega)/2+i\Delta(\Omega)]\tau}.
\label{eqn:Gbar}
\end{equation}
This exponential decay is important for the relaxation to the NESS irrespective of the initial state $\hat{\rho}_S$ of system $S$, while the naive perturbative calculation fails to capture this decay.
In van Hove's limit, we get the characteristic function of the NESS
\begin{widetext}
\begin{align}
\chi_\text{NESS}[\xi,\xi^*,\eta,\eta^*]
\simeq{}&
\exp\!\left(
-\frac{1}{2}
\xi^*
\xi
\frac{\Gamma^\beta(\Omega)}{\Gamma(\Omega)}
\right)
\nonumber\\
&{}\times
\exp\Biggl(
-\frac{1}{2}
\int_0^\infty\frac{d\omega}{2\pi}
\Gamma_{\eta\eta}^\beta(\omega)
+\lambda^2
\Re
\int_0^\infty\frac{d\omega}{2\pi}
\frac{
\Gamma_\eta^\beta(\omega)
}{\Omega-\omega+i0^+}
\int_0^\infty\frac{d\omega'}{2\pi}
\frac{
\Gamma_\eta^*(\omega')
}{\omega-\omega'-i0^+}
\nonumber\\
&\qquad\qquad\qquad\qquad\qquad\quad\ \ %
{}-\frac{1}{2}
\lambda^2
\frac{\Gamma^\beta(\Omega)}{\Gamma(\Omega)}
\int_0^\infty\frac{d\omega}{2\pi}
\frac{\Gamma_\eta(\omega)}{\Omega-\omega+i0^+}
\int_0^\infty\frac{d\omega'}{2\pi}
\frac{\Gamma_\eta^*(\omega')}{\Omega-\omega'-i0^+}
\Biggr)
\nonumber\displaybreak[0]\\
&{}\times
\exp\!\left[
\lambda
\Re
\int_0^\infty\frac{d\omega}{2\pi}
\left(
\Gamma_\eta^\beta(\omega)
-\Gamma_\eta(\omega)
\frac{\Gamma^\beta(\Omega)}{\Gamma(\Omega)}
\right)
\frac{1}{\Omega-\omega+i0^+}
\xi\right]
\label{eqn:CharFuncNESSWeak}
\end{align}
\end{widetext}
and the stationary current
\begin{gather}
\frac{1}{\lambda^2}\langle\hat{J}_\nu\rangle_\text{NESS}
\simeq
\sum_{\nu'=1}^M
\frac{
\Gamma^{(\nu)}(\Omega)
\Gamma^{(\nu')}(\Omega)
}{\Gamma(\Omega)}
[
f_\nu(\Omega)
-f_{\nu'}(\Omega)
]
\end{gather}
in the weak-coupling regime.
Note that in the weak-coupling regime the Fourier spectrum of the Green function (\ref{eqn:GFourier}) is approximated by $\tilde{G}(\omega)\simeq2\pi\delta(\omega-\Omega)$.

\subsection{Typical Pure NESS's}
Let us now turn our attention to typical pure NESS's. 
We provided the construction of typical pure NESS's $\ket{\phi}_\text{NESS}=\hat{W}\ket{\phi}$ 
in (\ref{eqn:PureNESSConst}), but it is practically easier to compute relevant quantities in the Heisenberg picture: instead of scattering a typical pure state $\ket{\phi}$, we scatter the observable $\hat{A}$ by the wave operator $\hat{W}$ as $\hat{W}^\dag\hat{A}\hat{W}$, and evaluate its expectation value in the initial typical pure state $\ket{\phi}$ in $\mathcal{H}_{S,\{E_\nu\}}$.

Let us consider the current operator $\hat{J}_\nu$ in (\ref{eqn:CurrentOpModel}).
On the basis of the solution (\ref{eqn:SolHeisenbergOps}) to the Heisenberg equations of motion, the expectation value of $\hat{J}_\nu$ in a typical pure NESS $\ket{\phi}_\text{NESS}$ is evaluated as follows.
As we stressed in the previous section, we start with finite-size reservoirs (e.g., in boxes with periodic boundary conditions) to sample a typical pure state $\ket{\phi}$ from $\mathcal{H}_{S,\{E_\nu\}}$ (from $\tilde{\mathcal{H}}_{S,\{E_\nu\}}$ in the bosonic case).
We compute the expectation value of $\hat{J}_\nu$ in the Heisenberg picture, take the thermodynamical (large-volume) limit $V\to\infty$, and then take the stationary limit $t\to\infty$:
\begin{widetext}
\begin{align}
&
\bras{\phi}{\text{NESS}}
\hat{J}_\nu
\ket{\phi}_\text{NESS}
\vphantom{\int}
\nonumber\\
&\qquad
=
\bra{\phi}
\hat{W}^\dag\hat{J}_\nu\hat{W}
\ket{\phi}
\nonumber\\
&\qquad
=
\lim_{t\to\infty}\lim_{V\to\infty}
2\Re\biggl(
-\lambda^2
\bar{G}^*(t)
(\bar{G}\circ\bar{K}^{(\nu)})(t)
\bra{\phi}
\hat{a}^\dag\hat{a}
\ket{\phi}
-i\lambda
\bar{G}^*(t)
\sum_{\bm{k}}
u_{\bm{k}\nu}^*
\bra{\phi}
\hat{a}^\dag
\hat{b}_{\bm{k}\nu}
\ket{\phi}
\nonumber\\
&\qquad\qquad\qquad\qquad\qquad
{}+i\lambda^3
\int_0^tdt'\,
[
\bar{G}^*(t)
(\bar{G}\circ\bar{K}^{(\nu)})(t')
+
(\bar{G}\circ\bar{K}^{(\nu)})^*(t)
\bar{G}(t')
]
\sum_{\nu'=1}^M\sum_{\bm{k}'}
u_{\bm{k}'\nu'}^*
e^{i(\omega_{k'\nu'}-\Omega)t'}
\bra{\phi}
\hat{a}^\dag
\hat{b}_{\bm{k}'\nu'}
\ket{\phi}
\nonumber\displaybreak[0]\\
&\qquad\qquad\qquad\qquad\qquad
{}+\lambda^2
\int_0^tdt'\,\bar{G}^*(t')
\sum_{\nu'=1}^M\sum_{\bm{k}}
\sum_{\bm{k}'}
u_{\bm{k}\nu}^*e^{-i(\omega_{k'\nu'}-\Omega)t'}u_{\bm{k}'\nu'}
\bra{\phi}
\hat{b}_{\bm{k}'\nu'}^\dag
\hat{b}_{\bm{k}\nu}
\ket{\phi}
\nonumber\displaybreak[0]\\
&\qquad\qquad\qquad\qquad\qquad
{}-\lambda^4
\int_0^tdt_1\int_0^tdt_2\,
\bar{G}^*(t_1)
(\bar{G}\circ\bar{K}^{(\nu)})(t_2)
\nonumber\\
&\qquad\qquad\qquad\qquad\qquad\qquad\qquad\qquad\qquad\ %
{}\times
\sum_{\nu'=1}^M\sum_{\nu''=1}^M
\sum_{\bm{k}'}\sum_{\bm{k}''}
u_{\bm{k}''\nu''}^*
e^{-i(\omega_{k'\nu'}-\Omega)t_1}
e^{i(\omega_{k''\nu''}-\Omega)t_2}
u_{\bm{k}'\nu'}
\bra{\phi}
\hat{b}_{\bm{k}'\nu'}^\dag
\hat{b}_{\bm{k}''\nu''}
\ket{\phi}
\biggr),
\label{eqn:TypicalNESSExp}
\end{align}
\end{widetext}
where $\bar{G}(t)$ is introduced in (\ref{eqn:Gbar}), and 
\begin{equation}
\bar{K}^{(\nu)}(t)
=\int_0^\infty\frac{d\omega}{2\pi}
\Gamma^{(\nu)}(\omega)
e^{-i(\omega-\Omega)t}
\end{equation}
with $\Gamma^{(\nu)}(\omega)$ defined in (\ref{eqn:GammaNu}).
Before the thermodynamical limit $V\to\infty$, the momenta $\bm{k}$ are discrete, and $\hat{b}_{\bm{k}\nu}$ as well as $u_{\bm{k}\nu}$ are scaled by $\sqrt{V}$ compared to those in the continuum limit, but by abuse of notation we use the same symbols for their discrete counterparts as those for continuous $\bm{k}$.

The problem is now reduced to the evaluation of the two-point expectation values of the canonical operators $\hat{a}$ and $\hat{b}_{\bm{k}\nu}$ in the typical pure state $\ket{\phi}$, i.e., $\bra{\phi}\hat{a}^\dag\hat{b}_{\bm{k}\nu}\ket{\phi}$ and $\bra{\phi}\hat{b}_{\bm{k}\nu}^\dag\hat{b}_{\bm{k}'\nu'}\ket{\phi}$.
Recall that the typical pure state $\ket{\phi}$ is sampled from $\mathcal{H}_{S,\{E_\nu\}}=\mathcal{H}_S\otimes(\mathop{\bigotimes}_{\nu=1}^M\mathcal{H}_{E_\nu}^{(\nu)})$ [from $\tilde{\mathcal{H}}_{S,\{E_\nu\}}=\tilde{\mathcal{H}}_S\otimes(\mathop{\bigotimes}_{\nu=1}^M\mathcal{H}_{E_\nu}^{(\nu)})$ in the bosonic case]. 
See (\ref{eqn:Hprod}) again.
In each reservoir $\nu$, the energy $E_\nu$ and the number of particles $N_\nu$ are distributed among different free modes $\hat{b}_{\bm{k}\nu}$ labelled by $\bm{k}$ and $\nu$.
Then, applying essentially the same argument as that for the canonical typicality \cite{ref:TasakiCanonicalTypicality,Lebowitz1,Popescu1-arXiv,*Popescu1}, the relevant modes are described by the grand canonical ensemble,
\begin{equation}
\bra{\phi}\hat{b}_{\bm{k}\nu}^\dag\hat{b}_{\bm{k}'\nu'}\ket{\phi}
\simeq
\langle
\hat{b}_{\bm{k}\nu}^\dag\hat{b}_{\bm{k}'\nu'}
\rangle_\text{gc}
=f_\nu(\omega_{k\nu})\delta_{\nu\nu'}\delta_{\bm{k}\bm{k}'}.
\end{equation}
As for the system operator $\hat{a}$, the pure state $\ket{\phi}$ is typically equivalent to the microcanonical state $\hat{\rho}_\text{mc}^{(S)}$ of the relevant Hilbert space $\mathcal{H}_S$ ($\tilde{\mathcal{H}}_S$ in the bosonic case), and we have
\begin{equation}
\bra{\phi}
\hat{a}^\dag\hat{b}_{\bm{k}\nu}
\ket{\phi}
\simeq
0,\quad
\bra{\phi}
\hat{a}^\dag\hat{a}
\ket{\phi}
\simeq
\langle
\hat{a}^\dag\hat{a}
\rangle_\text{gc},
\end{equation}
where $\langle{}\cdots{}\rangle_\text{gc}$ is evaluated in the state $\hat{\rho}_\text{gc}$ given in (\ref{eqn:GCmix}).
Substituting these results, the expectation value (\ref{eqn:TypicalNESSExp}) typically yields
\begin{widetext}
\begin{align}
\bras{\phi}{\text{NESS}}
\hat{J}_\nu
\ket{\phi}_\text{NESS}
\simeq
{-2}\lambda^2
\lim_{t\to\infty}\Re\biggl(&
\bar{G}^*(t)
(\bar{G}\circ\bar{K}^{(\nu)})(t)
\langle
\hat{a}^\dag\hat{a}
\rangle_\text{gc}
\nonumber\displaybreak[0]\\
&
{}-
\int_0^tdt'\,
\bar{G}^*(t')
\bar{K}^{(\nu)-}(t')
+\lambda^2
\int_0^tdt_1\int_0^tdt_2\,
\bar{G}^*(t_1)
\bar{K}^-(t_1-t_2)
(\bar{G}\circ\bar{K}^{(\nu)})(t_2)
\biggr),
\end{align}
\end{widetext}
where
\begin{subequations}
\begin{gather}
\bar{K}^-(t)
=\sum_{\nu=1}^M\bar{K}^{(\nu)-}(t),\\
\bar{K}^{(\nu)-}(t)
=\int_0^\infty\frac{d\omega}{2\pi}
f_\nu(\omega)
\Gamma^{(\nu)}(\omega)
e^{-i(\omega-\Omega)t}.
\end{gather}
\end{subequations}
In the long-time limit $t\to\infty$, the term containing $\langle\hat{a}^\dag\hat{a}\rangle_\text{gc}$ decays out and the above typical expectation value yields the steady current (\ref{eqn:NESSCurrent}) in the NESS\@.

\section{Summary}
\label{sec:Summary}
It is a priori unclear whether a single pure state can represent an ensemble describing a nonequilibrium state.    
In this article, we have shown that it is actually the case for NESS's, even in the presence of a small quantum system coupled to reservoirs.
We have identified a relevant Hilbert space $\mathcal{H}_\text{NESS}$ whose typical pure states $\ket{\phi}_\text{NESS}$ are essentially equivalent to a NESS of the system: the typicality holds also for NESS's.
To this end, we have stressed that the irreversibility of the relaxation to a unique NESS is important and nonperturbative treatment is required.

Interesting future subjects include the issue whether fluctuation theorems \cite{Gaspard1,Saito1,ref:SaitoTasaki-ExtendedClausius,ref:FT-RMP-EspositoMukamel,ref:FT-RMP-CampisiHanggi} hold for \textit{isolated} quantum systems described by typical \textit{pure} states (cf.\ \cite{Monnai1}).
This further leads us to better understanding on the physics of nonequilibrium systems and processes.

\acknowledgments
This work is partially supported by the Top Global University Project from the Ministry of Education, Culture, Sports, Science and Technology (MEXT), Japan, by Grants-in-Aid for Young Scientists (B) (No.\ 26800206) and for Scientific Research (C) (No.\ 26400406) from the Japan Society for the Promotion of Science (JSPS), and by a Waseda University Grant for Special Research Projects (No.\ 2015K-202).


%

\end{document}